\documentclass[aps,showpacs]{revtex4}
\usepackage{epsf,latexsym}

\begin{document}

\title{Mass and temperature limits for blackbody radiation}
%\author{Alessandro Pesci\footnotetext{e-mail: pesci@bo.infn.it}}
\author{Alessandro Pesci}
\email{pesci@bo.infn.it}
\affiliation
{INFN, Sezione di Bologna Via Irnerio 46, I-40126 Bologna, Italy}

\begin{abstract}
A spherically symmetric distribution
of classical blackbody radiation
is considered, at conditions in which
gravitational self-interaction effects become not negligible.
Static solutions to Einstein field equations are searched
for, for each choice of the assumed central energy density.
Spherical cavities at thermodynamic equilibrium, 
i.e. filled with blackbody radiation,
are then studied,
in particular for what concerns
the relation among the mass $M$ of the 
ball of radiation contained in them
and their temperature at center and at the boundary.
For these cavities it is shown, in particular, that:
i) there is no absolute limit to $M$
as well to their central and boundary temperatures;  
ii) when radius $R$ is fixed, however,
limits exist both for mass 
and 
for boundary energy density $\rho_B$:
$M \leq  K \ M_{S}(R)$ and
$\rho_B \leq Q/R^2$,
with $K = 0.493$ and $Q = 0.02718$, dimensionless,
and $M_S(R)$ the Schwarzschild mass for that radius.
Some implications of the existence and the magnitude
of these limits are considered.  
Finally 
the radial profiles for entropy for these systems
are studied,
in their dependence on the mass (or central temperature)
of the ball of radiation.
\end{abstract}


\maketitle

\section{Introduction}
The original motivation for this work
is to explore the relation
between entropy and gravity in a classical
context (i.e. not quantum).
The goal is to see if, 
under strong gravitational
interaction circumstances, some properties
peculiar to black hole entropy (such as for example
its area-scaling law), apparent in a semiclassical 
approach, 
can also be retrieved for conventional entropy
in a purely classical scenario, 
in which the fields too, in addition to the background, are classical.
Having this in mind, systems consisting 
entirely of blackbody radiation
are chosen for investigation,
with the hope that for them, both entropy can be readily
calculated and, also, strong gravitational interaction circumstances
can be reached.
During the analysis of such systems 
we have found for them
the existence of peculiar mass and temperature limits
for each assigned radius $R$.

The idea that under strong gravitational interaction
conditions conventional entropy can acquire properties
similar to black hole entropy is not new.
In particular in \cite{Oppen1} it has been shown that
the entropy of a spherically symmetric self-gravitating system
in equilibrium becomes area scaling as the radius approaches
the Schwarzschild value; in this limit the total entropy
of the body is accounted for only by the entropy of the more external
spherical shells.  
In \cite{Oppen2} the entropy of a spherically symmetric system
in equilibrium consisting of a perfect fluid with constant
density has been explicitly calculated. The entropy has been found
to loose its extensivity property and to reside more and more in the
external spherical shells as the gravitational interaction
goes stronger.
The limit $R = 2M$ 
($M$ is the mass, or mass-energy, as spelled out in \cite{Wheeler},
known also as ADM mass)
cannot be reached in this case, 
as equilibrium is possible only
if $R \geq (9/4)M$ (see, for instance, \cite{Wheeler}).
One would like to explicitly calculate entropy for equilibrium systems
with other equations of state, possibly going to explore also the limit
$R \simeq 2M$. But it turns out that for all viable examples
the radius of the system becomes infinite. 
This is a problem because in this case it becomes unclear what is intended
with the concepts
of ``more external shells'' and ``mass'' of the system. 
The recourse to different
equations of state for different radii can give realistic models
with finite size, but can hide the relation between entropy and gravity
behind the circumstantial
peculiarities of the constructed model.

As well known (see for example \cite{Wheeler}) and we will recall later,
a gravitationally self-interacting static distribution 
of blackbody radiation
is in general precisely one of these infinite size systems. 
From a thermodynamical point of view this system
is particularly appealing due to its simplicity,
as its state is locally determined by only one 
thermodynamical observable,
for example its temperature $T$.
It is then tempting to use such a system
to investigate
the connection between entropy and gravity 
and
we can imagine to circumvent the problem of its being of infinite size
by forcing it to be adiabatically confined inside a finite size cavity.
By applying a sufficiently high compression
one could then expect the system to
go towards its Schwarzschild limit.
In principle it becomes then possible
to study the behaviour of a pure classical system, 
this cavity filled with blackbody radiation
(in our study all what is needed of blackbody radiation are
purely classical properties, as we will see),
in a strong gravitational regime.
At these conditions, 
by the way, 
some quantum behaviour is expected to come on
for this system,
as, of course, pair production or other effects \cite{Hernandez}.
We decide however to study the system
at a purely classical level, 
and to postpone
any quantum consideration (pair production) 
at the end, once the characteristics
classically expected
are known.
All this is the content of the following section.

\section{Field equations and their solutions}
We consider a perfect fluid consisting of a photon gas
at thermodynamic equilibrium
(somehow along the lines of \cite{photonstars}, 
in which this same system is appoached,
even if with different context and motivations)
or, without quantum terminology, consisting of blackbody
radiation.
Its equation of state is

\begin{eqnarray}\label{state}
p = \rho/3,
\end{eqnarray}
where $\rho$ and $p$ are respectively the energy density
and pressure in the rest frame of the fluid element.
Its stress-energy tensor is

\begin{eqnarray}\label{stress-energy}
T_{\alpha\beta} = (\rho + p) u_{\alpha} u_{\beta} + p g_{\alpha\beta},
\end{eqnarray}
where $u_{\alpha}$ is the fluid element 4-velocity and
$g_{\alpha\beta}$ is the metric tensor.
In the case of static equilibrium configurations
(necessarily with spherical symmetry),
from (\ref{state}) and (\ref{stress-energy})
Einstein field equations in Schwarzschild coordinates
read 

\begin{eqnarray}\label{einstein1}
\nonumber
& & \frac{h^\prime}{r h^{2}} + \frac{h-1}{h r^{2}} = 8 \pi \rho \ \ \ \ 
\qquad \qquad \qquad \qquad \quad \qquad \qquad (tt) \\
& & \frac{f^\prime}{r f h} - \frac{h-1}{h r^{2}} = \frac{8 \pi}{3} \rho \ \ \ 
\qquad \qquad \qquad \qquad \quad \qquad \qquad (rr) \\
\nonumber
& & \frac{f^\prime}{r f h} - \frac{h^\prime}{r h^{2}} +
\frac{1}{\sqrt{f h}} \left( \frac{f^\prime}{\sqrt{f h}} \right)^\prime = 
\frac{16 \pi}{3} \rho \ \ \  \qquad \qquad (\theta\theta)=(\phi\phi),
\end{eqnarray}
where the derivative symbol means derivative
with respect to radial coordinate (area radius) $r$ and
where the metric is

\begin{eqnarray}\label{metric}
ds^2 = -f(r) \ dt^2 + h(r) \ dr^2 + r^2 \ d\Omega^2,
\end{eqnarray}
with $f$ and $h$ two non-negative functions of $r$, to be determined.

For static gravitational fields, Tolman relation \cite{Tolman1930}
prescribes that $T \sqrt{-g_{00}}$ be constant throughout the system,
where $T$ is the temperature in the rest frame of the fluid element.
This means for us that 

\begin{eqnarray}\label{tolman}
T(r) \sqrt{f(r)} = const.
\end{eqnarray}
For blackbody radiation, Stefan-Boltzmann classical result
reads (in conventional units)

\begin{eqnarray}\label{stefan-boltzmann}
\rho = \frac{4}{c} \sigma T^4,
\end{eqnarray}
with $c$ the speed of light in vacuum and $\sigma$
the Stefan-Boltzmann constant.
Applying this equation locally throughout
our system (equivalence principle),
from equation (\ref{tolman}) we have thus

\begin{eqnarray}\label{C}
\rho(r) f^2(r) = const \equiv \frac{C}{8 \pi},
\end{eqnarray}
with $C$ some constant with the dimensions
of energy density (length$^{-2}$ in geometrized units).
Note that while the value of $C$ depends
on the chosen unit for length,
the quantities $\rho/C$ and $r \sqrt{C}$ are dimensionless.

Making use of (\ref{C}), 
the first and the second of equations (\ref{einstein1}) become

\begin{eqnarray}\label{einstein2_1}
& & \frac{C}{f^2} = \frac{h^\prime}{r h^2} + \frac{h-1}{h r^2} \\
\label{einstein2_2}
& & \frac{C}{3 f^2} = \frac{f^\prime}{r f h} - \frac{h-1}{h r^2},
\end{eqnarray}
while the third, as can be verified
(and as noticed in \cite{photonstars}), 
is always identically satisfied
if equations (\ref{einstein2_1}-\ref{einstein2_2}) are satisfied,
as a confirmation of Tolman result.

Before we start to solve these coupled differential equations,
let us study some general characteristics which 
the solutions $f(r)$ and $h(r)$
should satisfy. 
$h(r)$ can be put in the form

\begin{eqnarray}\label{hm}
h(r) = \frac{1}{1-2 m(r)/r},
\end {eqnarray}
(see for example \cite{Wheeler}) with $m(r)$
given by

\begin{eqnarray}\label{ADM}
m(r) = \int_0^r{4 \pi {\tilde r}^2 \rho({\tilde r}) d{\tilde r}} =
\frac{C}{2} \int_0^r{\frac{{\tilde r}^2}{f^2({\tilde r})} d{\tilde r}}, 
\end{eqnarray}
that is, $m(r)$ is the ADM mass inside $r$. 

For each allowed equilibrium configuration,
in the limit $r \rightarrow 0$

\begin{eqnarray}\label{mnear0} 
m(r) \simeq \frac{4}{3} \pi r^3 \rho(0),
\end{eqnarray}
so that $m(r)/r \rightarrow 0$. This implies $h(0) = 1$ for each solution
with $\rho(0)$ finite.
As regards $f$, from (\ref{C}) we have

\begin{eqnarray}\label{fbegin}
\rho(0) f^2(0) = \frac{C}{8 \pi},
\end{eqnarray}
so that finite $\rho(0)$ implies a non zero $f(0)$.   
Taking the derivative of (\ref{hm}) and using (\ref{mnear0}) and (\ref{fbegin}),
we have that in the limit $r \rightarrow 0$, $h^\prime(r) \simeq \frac{16}{3} \pi \rho(0) r$
and $f^\prime(r) \simeq r f(0) \frac{16}{3} \pi \rho(0)$,
so that $h^\prime(0) = 0$ and $f^\prime(0) = 0$.

We want now to solve equations (\ref{einstein2_1}-\ref{einstein2_2}).
We can write equation (\ref{einstein2_2}) (as in \cite{photonstars}) as

\begin{eqnarray}\label{h15}
h = \frac{3 f (f + r f^\prime)}{C r^2 + 3 f^2}
\end{eqnarray} 
and, substituting this expression of $h$ in equation (\ref{einstein2_1}),
we obtain 

\begin{eqnarray}\label{fderivata2}
f^{\prime\prime} = \frac{6 C r f^2 + 6 C r^2 f f^\prime - 6 f^3 f^\prime + 2 C r^3 {f^\prime}^2}
{C r^3 f + 3 r f^3}
\end{eqnarray}
so that equations (\ref{h15}-\ref{fderivata2}) are equivalent 
to the starting equations (\ref{einstein2_1}-\ref{einstein2_2}).
Using, in addition, the previous results about the behaviour
of $f$ near the origin, 
the problem is thus reduced to find the solutions $f$
of the second order equation (\ref{fderivata2})
with initial conditions

\begin{eqnarray}\label{condiniz}
& f(0) = \sqrt{\frac{C}{8 \pi \rho(0)}}; & \qquad \qquad f^\prime(0) = 0.
\end{eqnarray}

For each choice of the pair $C$, $\rho(0)$
we have a solution $f(r)$.
Note however from equation (\ref{C}) that
physical configurations $\rho = \rho(r)$ 
depend only on the ratio $C/f^2$ and not on
$C$ and $f$ separately. 
This implies that for any chosen $\rho(0)$,
the (infinite number of) solutions $f$ that we find
when $C$ is varied, correspond always to the same physical
configuration $\rho = \rho(r)$.
To fix $\rho(0)$ corresponds to uniquely fix
the physical configuration, irrespective of the separate values
of $C$ and $f$. 
Without loss of generality we can then study
the solutions to equation (\ref{fderivata2}) for
a single arbitrary choice of the value of $C$ ($\not= 0$);
let us put for simplicity $C = 1$.
All possible static configurations of our system
are then in 1-1 correspondence with all the solutions
$f$ of equation

\begin{eqnarray}\label{fderivata2bis}
f^{\prime\prime} = \frac{6 r f^2 + 6 r^2 f f^\prime - 6 f^3 f^\prime + 2 r^3 {f^\prime}^2}
{r^3 f + 3 r f^3}
\end{eqnarray}
with initial conditions

\begin{eqnarray}\label{condinizbis}
& f(0) = \sqrt{\frac{1}{8 \pi \rho(0)}}; & \qquad \qquad f^\prime(0) = 0,
\end{eqnarray}
where $\rho(0)$ is any positive number
and with $h$ given by (\ref{h15}).
Due to (\ref{stefan-boltzmann}), each specification of $\rho(0)$
is equivalent to a specification of the central temperature $T_C$.

Some symmetry considerations can now help to understand
how the set of solutions $f$ (and $h$) is arranged.
Consider the transformation

\begin{eqnarray}\label{transformationr}
& & r \rightarrow {\tilde r} = \lambda r \\
\label{transformationf}
& & f \rightarrow {\tilde f} = \lambda f .
\end{eqnarray}
We have

\begin{eqnarray}
\nonumber
{\tilde f}({\tilde r}) = \lambda f({\tilde r}/\lambda)
\end{eqnarray}
or

\begin{eqnarray}
\nonumber
{\tilde f}(r) = \lambda f(r/\lambda)
\end{eqnarray}
and then

\begin{eqnarray}
\nonumber
{\tilde f}^\prime(r) = f^\prime(r/\lambda)
\end{eqnarray}
and

\begin{eqnarray}
\nonumber
{\tilde f}^{\prime\prime}(r) = \frac{1}{\lambda} f^{\prime\prime}(r/\lambda)
\end{eqnarray}
so that, if $f(r)$ is a given solution, ${\tilde f}(r)$ is also a solution,
as can be easily verified going through equation (\ref{fderivata2bis}).
${\tilde f}(r)$ is such that ${\tilde f}(0) = \lambda f(0)$.
On the other hand we know that, 
starting from a given solution $f_1$,
we can obtain all solutions 
if we change the initial conditions,
$f_1(0) \rightarrow f_{\lambda}(0) = \lambda f_1(0)$,
with $\lambda$ spanning all positive real values.
Thus,
in terms of an assigned solution $f_1(r)$,
a function $f_\lambda$ is a solution if and only if

\begin{eqnarray}\label{allsolutionsf}
f_\lambda(r) = \lambda f_1(r/\lambda)
\end{eqnarray}
with $\lambda > 0$.
For the solutions $h$ we obtain

\begin{eqnarray}\label{allsolutionsh}
h_\lambda(r) = h_1(r/\lambda),
\end{eqnarray}
where $h_\lambda(r)$ and $h_1(r)$ corresponds to 
$f_\lambda(r)$ and $f_1(r)$
through (\ref{h15}).

Summing up, once we know the behaviour of 
solutions $f$ and $h$ for a given particular initial value
$f(0) = a_1$ of $f$, arbitrarily fixed ($>0$),
we can readily get the corresponding behaviours
of the solutions for every other assigned initial value $a_2$.
Let us study for simplicity the case $f(0) = 1$,
that is the solutions to (\ref{fderivata2bis})
with initial conditions

\begin{eqnarray}\label{condinizter}
& f(0) = 1; & \qquad \qquad f^\prime(0) = 0.
\end{eqnarray}

\begin{figure}\leavevmode
\begin{center}
\epsfxsize=8cm
\epsfbox{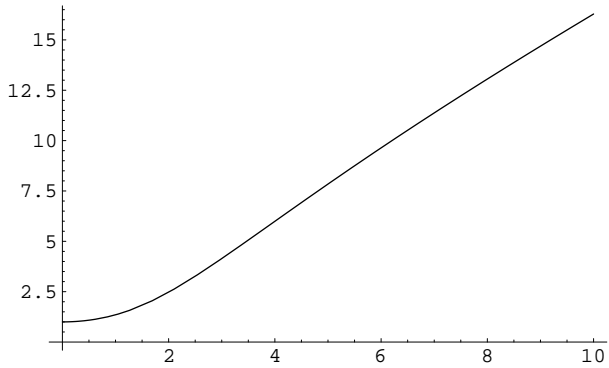}
\caption{$f$ versus $r$.}
\label{f}
\end{center}
\end{figure}

\begin{figure}\leavevmode
\begin{center}
\epsfxsize=8cm
\epsfbox{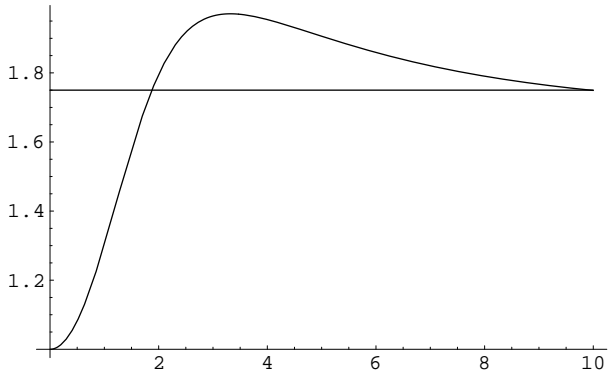}
\caption{$h$ versus $r$.}
\label{h}
\end{center}
\end{figure}

From numerical integration of equation (\ref{fderivata2bis})
and from equation (\ref{h15}) we obtain
the solutions $f$ and $h$ reported in Figures \ref{f} and \ref{h}. 
%%%%%%%%%%%%%%%%%%%%%%%%%%%% figure f %%%%%%%%%%%%%%%%%
%%%%%%%%%%%%%%%%%%%%%%%%%%%%%%%%%%%%%%%%%%%%%%%%%%%%%%%
%%%%%%%%%%%%%%%%%%%%%%%%%%%% figure h %%%%%%%%%%%%%%%%%
%%%%%%%%%%%%%%%%%%%%%%%%%%%%%%%%%%%%%%%%%%%%%%%%%%%%%%%
The structure of the corresponding equilibrium configuration
can be seen in Figures \ref{rho} and \ref{2m_over_r},
that respectively report $\rho$
and $2m/r$, as a function of $r$.
Our choice $C = 1$ implies that $\rho$ and $r$ are given here
respectively in units of $C$ and $C^{-1/2}$.
From (\ref{C}) and (\ref{allsolutionsf}) the properties
of $\rho$ for every other solution can readily be inferred.
Note that $\rho(r)$ approaches asymptotically 0 as $r \rightarrow \infty$.
This implies that for this configuration and, 
by scaling property (\ref{allsolutionsf}),
for every other solution, the radius of the system is infinite.

Let us consider in some detail the functions 
$h$  and $2m/r$ (Figures \ref{h} and \ref{2m_over_r}).
$h$ starts from 1 at $r=0$, 
and $2m/r$ from 0;  
they reach their maxima,
after which they go
towards their own limiting values \cite{photonstars},

\begin{eqnarray}\label{limith}
h(\infty) = 7/4
\end{eqnarray}
and

\begin{eqnarray}\label{limit}
\left(\frac{2m}{r}\right)_\infty = 3/7,
\end{eqnarray}
shown in the Figures, 
with damped oscillations around them,
in which both the amplitudes
of the deviations from the limiting values as well their derivatives
at any order reduce asymptotically to 0.
Analytically both limiting values can easily be verified.
In fact consider for example that from (\ref{hm}) we have

\begin{eqnarray}\label{2m_over_r_formula}
\frac{2m}{r} = \frac{h-1}{h}.
\end{eqnarray}
In equation (\ref{einstein2_1}) we see then that,
at extremal points,
that is when $r$ has values $r^*$ such that
$h^\prime(r^*) = 0$ (that is $(2m(r)/r)^\prime_{r=r^*} = 0$),
we have

\begin{eqnarray}\label{extrema}
\frac{2m(r^*)}{r^*} = \frac{{r^*}^2}{f^2(r^*)}. 
\end{eqnarray}
As extremal values tend to $(2m/r)_\infty$ 
when $r \rightarrow \infty$,
from equation (\ref{einstein2_2}) considered in this limit,
using (\ref{2m_over_r_formula}) we have

\begin{eqnarray}
\nonumber
\frac{1}{3} \left(\frac{2m}{r}\right)_\infty =
1 - \left(\frac{2m}{r}\right)_\infty - \left(\frac{2m}{r}\right)_\infty =
1 - 2  \left(\frac{2m}{r}\right)_\infty
\end{eqnarray} 
and thus equation (\ref{limit}),
as well
the result (\ref{limith}).
Here we made use of the fact that
in the limit $r \rightarrow \infty$,
$(r^2/f^2)^\prime = 0$ implies $f^\prime = f/r$.
Note that equations (\ref{2m_over_r_formula}) 
and (\ref{allsolutionsh})
determine 
$2m/r$ for every solution; we see that $2m/r$ scales
as $h$. 
From scaling property (\ref{allsolutionsh}) and 
from equations (\ref{condinizbis}) and ({\ref{transformationf}),
the asymptotic behaviour of $h$ and $2m/r$ when $r \rightarrow \infty$
implies that in the limit of very large central density $\rho(0)$
or very large $T_C$, $h(r) \simeq const = 7/4$ 
and $2m(r)/r \simeq const = 3/7$,
for every $r$,
except the region of $r$ very near to 0, in which $h$ goes from 1
to 7/4 
and $2m/r$ goes from 0 to 3/7,
with damped oscillations.

As regards $f$,
from (\ref{einstein2_2}) we see that $f^\prime(r) \geq 0, \forall r$
and from (\ref{extrema}) we have that in the limit 
$r \rightarrow \infty$, $f(r) \simeq \sqrt{7/3} \ r$ (see \cite{photonstars}).
Due to equation (\ref{allsolutionsf}) this implies

\begin{eqnarray}
\nonumber
\lim_{\lambda\to0} f_{\lambda}(r) =  
\lim_{\rho(0)\to\infty} f_{\lambda}(r) =
\lim_{T_C\to\infty} f_{\lambda}(r) = 
\lim_{\lambda\to0} \lambda f(r/\lambda) =
\lim_{\lambda\to0} \ \lambda \ \frac{7}{3} \ \frac{r}{\lambda} =
\sqrt{\frac{7}{3}} \ r, \qquad \qquad (r fixed)
\end{eqnarray}
and then

\begin{eqnarray}\label{limitT}
\lim_{\rho(0)\to\infty} \rho(r) =
\lim_{T_C\to\infty} \rho(r) =
\lim_{T_C\to\infty} \frac{1}{8 \pi f_{\lambda}^2(r)} = 
\frac{3}{56 \pi} \frac{1}{r^2}
\qquad \qquad (r fixed)
\end{eqnarray}
that is when the central density (central temperature)
is allowed to increase without limit, the density (temperature)
at each fixed radius $r$ does not increase unlimitedly
but approaches a limiting value proportional to 
$r^{-2}$ ($r^{-1/2}$).

Finally, from scaling properties
(\ref{allsolutionsf}-\ref{allsolutionsh}),
from initial conditions (\ref{condinizbis})
and from the behaviour of $f(r)$, $h(r)$ and $\rho(r)$
near $r=0$, shown in Figures \ref{f}, \ref{h} and \ref{rho},
we have that in the limit of very low central density,
$f(r)$, $h(r)$ and $\rho(r)$ are constant
in a ever increasing range of $r$, 
so that the flat-space limit for constant low densities
is recovered.

%%%%%%%%%%%%%%%%%%%%%%%%%%%% figure rho %%%%%%%%%%%%%%%%%
\begin{figure}\leavevmode
\begin{center}
\epsfxsize=8cm
\epsfbox{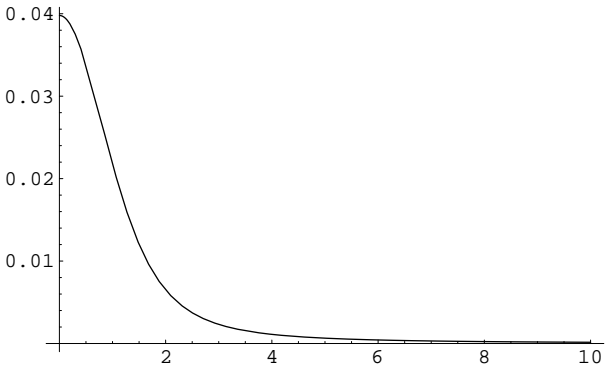}
\caption{$\rho$ versus $r$.}
\label{rho}
\end{center}
\end{figure}
%%%%%%%%%%%%%%%%%%%%%%%%%%%%%%%%%%%%%%%%%%%%%%%%%%%%%%%
%%%%%%%%%%%%%%%%%%%%%%%%%%%% figure 2m_over_r %%%%%%%%%%%%%%%%%
\begin{figure}\leavevmode
\begin{center}
\epsfxsize=8cm
\epsfbox{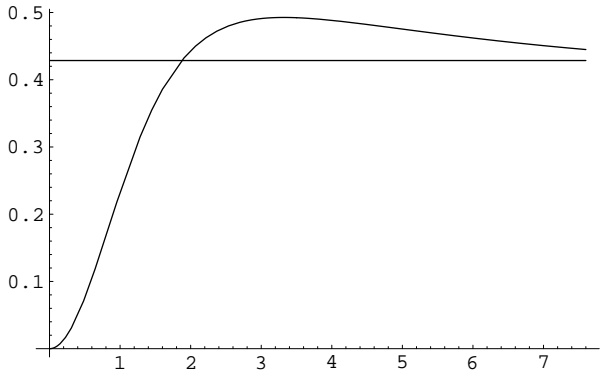}
\caption{$2m/r$ versus $r$.}
\label{2m_over_r}
\end{center}
\end{figure}
%%%%%%%%%%%%%%%%%%%%%%%%%%%%%%%%%%%%%%%%%%%%%%%%%%%%%%%

To obtain a system with finite size 
we can imagine
to enclose our radiation
inside a perfectly insulating spherical cavity
with area radius $R$,
with its center coinciding
with the center of the radiation distribution
and to assume that thermodynamical equilibrium
is reached, that is
the temperature of the walls is equal to
the temperature of radiation at $r = R$.
This will correspond to a certain central
energy density $\rho_C$ for the cavity. 
Due to the local uniqueness of solutions
to equation (\ref{fderivata2bis}) with initial conditions
(\ref{condinizbis}), 
this is equivalent to take a certain (physical) solution
to equation (\ref{fderivata2bis})
(that with $\rho(0) = \rho_C$),
to cancel all radiation outside $r = R$ and to put
an insulating spherical wall (of negligible thickness and mass)
at $r = R$ exactly at the temperature
the radiation has there.
As regards energy density radial profile $\rho(r)$,
this means that there is a 1-1 correspondence between
all admitted cavities 
filled with blackbody radiation 
(that is every admitted central temperatures or
boundary temperatures as well radii)
and
all possible $\rho(r)$ distributions 
in finite ranges [0,$R$], 
where $\rho(r) = 1/(8 \pi f^2(r))$ and $f$ is solution
to (\ref{fderivata2bis}) with initial conditions (\ref{condinizbis}).

Let us assume now, 
as allowed by Birkhoff theorem \cite{Wheeler}, \cite{Birkhoff}, 
that the Schwarzschild coordinates
be chosen in such a way that outside the cavity
exactly the Schwarzschild form of the metric obtains

\begin{eqnarray}\label{schwarzschild}
ds^2 = - ( 1 - 2M/r) \ dt^2 + \frac{1}{1-2M/r} \ dr^2 
+ r^2 d\Omega^2.
\end{eqnarray}
The parameter $M$ here is precisely the mass-energy or ADM mass 
as given by (\ref{ADM}) with $r = R$, that is

\begin{eqnarray}\label{cavityADM}
M = \int_0^R{4 \pi r^2 \rho(r) dr} =
\frac{C}{2} \int_0^R{\frac{r^2}{f^2(r)} dr}, 
\end{eqnarray}
This mass parameter $M$
is the mass probed
by Kepler-like experiments with gravitating
bodies far away from the system.
In our context
it thus represents meaningfully
the mass of the ball of radiation
contained in the cavity.
As the metric outside has been explicitly chosen,
the values of the metric tensor components $g_{\alpha\beta}$
at the boundary are now fixed.
Going at this boudary from inside,
continuity of metric tensor components
at the boundary of the cavity
requires that 

\begin{eqnarray}\label{boundaryf}
& & f(R) = 1 - 2M/R \\
\label{boundaryh}
& & h(R) = 1/(1-2M/R).
\end{eqnarray}
Equations (\ref{hm}),(\ref{ADM}) and (\ref{cavityADM}) show
that $h$ fulfills condition (\ref{boundaryh}).
Making use of the freedom we have to choice $C$ and $f$
separately, provided the ratio $C/f^2$ be held fixed,
we can always find a value of $C$ such that the corresponding
$f$ fulfills the boundary condition (\ref{boundaryf}). 
In other words, once a single coordinate system is explicitly chosen,
the arbitrariness in the choice of
the constant $C$ in equations 
(\ref{einstein2_1})-(\ref{einstein2_2}) is lost.
In the case of a cavity filled with radiation,
with central density $\rho_C$,
condition (\ref{boundaryf}) requires $C$ to be

\begin{eqnarray}\label{cavityC}
C = \frac{f^2(R)}{f_1^2(R)} =
\frac{(1-2M/R)^2}{f_1^2(R)},
\end{eqnarray}
where with $f_1$ we intend here the solution
$f$ for the case $C = 1$ and $\rho(0) = \rho_C$. 
Each choice of central density $\rho(0)$
fixes the mass $M$ within a given radius $R$ and then 
separately $C$ and $f$.

From the results obtained so far some
consequences can be drawn.

A first consequence
regards mass. We can infer it
simply from
the behaviour of the functions $2m(r)/r$
for equilibrium configurations,
as detailed above.
The non-vanishing asymptotic limit (\ref{limit})
implies that there is no limit on the
mass $M$; 
we have however that for each fixed $R$

\begin{eqnarray}\label{mass-limit}
\frac{2M(R)}{R} \leq K = 0.493,
\end{eqnarray}
with $K$ dimensionless not depending on $R$. 
The value of $K$ can be read
directly in Fig. \ref{2m_over_r} as the value
of the maximum of the function $2m/r$ there reported.
This means that, whichever is the energy density
or the temperature of the blackbody radiation
at equilibrium inside a cavity of radius $R$,
its mass can never exceed a fraction $K$ of
the Schwarzschild mass $M_S = R/2$ for that radius.
Let consider a cavity at its mass limit $M_{limit}$,
for concreteness the configuration described
in Fig. \ref{2m_over_r} with area radius $R$ equal
to the extremal point. 
Suppose to supply it, in some unspecified manner, 
with some positive amount 
of energy $\Delta E$, while mantaining
fixed its radius.
What we should obtain at equilibrium, if reachable, is
simply a cavity with new conditions
at the center and at the boundary
and, by conservation of energy, with a mass
$M = M_{limit} + \Delta E$;
but we already know that for the given radius
an equilibrium configuration with such a mass value
does not exist. We have therefore that no additional
mass-energy can be stored at equilibrium in that given region
of space if, in the absorption process, it is put
in the form of blackbody radiation.
For masses lower than the limit (\ref{mass-limit}),
at least one equilibrium configuration exists,
but also cases with 2, 3, 4, ... equilibrium
configurations are allowed with different central temperatures;
this can be readily understood
taking into account
the properties of the function $2m/r$ described above.
Fig. \ref{mass_vs_T} permits to see, for a cavity
with fixed area radius $R$ at equilibrium, 
the dependence of its mass on the
central density or temperature.
It reports $2M/R$ as a function
of $ z \equiv \rho(0) R^2$.
%%%%%%%%%%%%%%%%%%%%%%%%%%%% figure mass_vs_T %%%%%%%%%%%%%%%%%
\begin{figure}\leavevmode
\begin{center}
\epsfxsize=8cm
\epsfbox{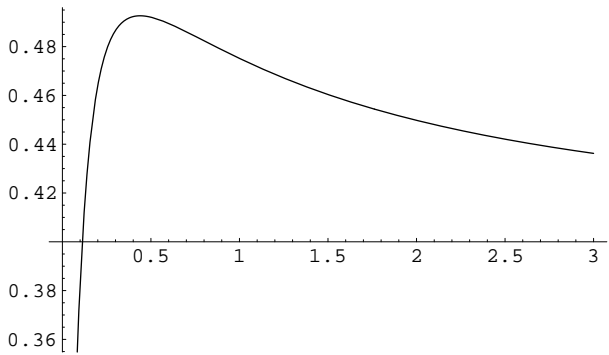}
\caption{$2M/R$ versus $z \equiv \rho(0) R^2$ ($R$ is fixed).}
\label{mass_vs_T}
\end{center}
\end{figure}
%%%%%%%%%%%%%%%%%%%%%%%%%%%%%%%%%%%%%%%%%%%%%%%%%%%%%%%
The plot is done in terms of $z$ instead of $\rho$
in order to obtain for $2M/R$ an universal shape, 
i.e. not depending on $R$.
The expression for $z$
indicates how to rescale the abscissa 
to obtain, from the given plot,
the actual function $2M(\rho(0))/R$
for each fixed value of $R$.
The explicit form of the plotted function, 
expressed in terms of the solution
$h_{f(0)=1}$ for initial conditions (\ref{condinizter})
(the solution $h$ presented in Fig. \ref{h}), is

\begin{eqnarray}\label{mass-temperature}
\frac{2M(z)}{R} = 
\frac{h_{f(0)=1}(\sqrt{8 \pi z}) - 1}{h_{f(0)=1}(\sqrt{8 \pi z})}
\end{eqnarray}  
as can be easily derived from equation (\ref{2m_over_r_formula}) and
from the scaling property (\ref{allsolutionsh}).
In the Figure the maximum is, obviously, $(2M/R)_{max} = 0.493$
and the extremal point ${\hat z}$, is ${\hat z} = 0.439$;
this values can be used to calculate various quantities
for the extremal equilibrium configuration.
If, for example, the radius is the solar radius,
$R = R_{\odot}$, we can calculate the extremal values of mass
$M_{max}$ and central density $\rho(0)_{extr}$ as follows:
$M_{max} = (2M/R)_{max} \ R_{\odot}/2 = 1.16 \cdot 10^5 \ M_{\odot}$,
with $M_{\odot}$ the solar mass, and
$\rho(0)_{extr} = {\hat z}/R_{\odot}^2 = 0.91 \cdot 10^{-22}$ cm$^{-2}$,
to which corresponds, through (\ref{stefan-boltzmann}), a temperature 
$T_{extr} = 1.95 \cdot 10^{10}$K.
Note that this temperature gives $k T_{extr} = 1.68$ MeV, with $k$
the Boltzmann constant. This implies that in the core of this system
many photons are well above threshold for pair production and  
our assumption of a fluid consisting of blackbody radiation alone
is no longer viable. Our results reside on two characteristics of
blackbody radiation systems, namely the equation of state
(\ref{state}) (and many systems besides blackbody radiation
obey it, for example ideal Fermi gases in the extreme relativistic
limit) and the peculiar energy-temperature relation (\ref{stefan-boltzmann}).
Only systems for which both these properties hold true,
are describable in the way we have shown and 
the eventual existence of conditions
could be investigated,
for which this is
the case for a mixture of blackbody photons and electrons
and positrons at equilibrium through the reaction 
$\gamma \gamma \rightleftharpoons e^+ e^-$.
On the other hand, as $T_{extr} \propto R^{-1/2}$,
when the radius is
of the order of $\sim 100 \ R_{\odot}$ or larger
the contribution of pair production starts to be
highly depressed and thus in any case
our assumptions and results 
in principle permit to describe faithfully the system,
not only at low densities but even up to its mass limit.
As the mass limit (\ref{mass-limit}) is very high,
systems of this size can store in this manner
mass energies of orders of $10^7 M_{\odot}$
while remaining at equilibrium. 
Quite interestingly 
we can have thus physically allowed 
and easily (classically) describable systems,
which are extremely compact,
with $10^7 M_{\odot}$ masses
within a size of few hundreds of light travel seconds
(i.e. a fraction of AU). 

In Fig. \ref{mass_vs_T} we see that
once the extremal central energy density or temperature
are overcome, equilibrium configurations are still possible,
at lower values of $M$. We have thus larger central temperatures,
which give rise however to a lower total mass for the system. 
This is by no way a problem for energy conservation, obviously.
The total energy of the system, given by $M$,
contains in fact both the component of proper energy $\rho$
and the (negative) component of gravitational potential 
energy \cite{Wheeler}, \cite{Misner}.
Evidently, when the extremal temperature is overcome,
the system settles down into a new equilibrium configuration
with a stronger gravitational energy component.

A second consequence regards energy density or temperature. 
Note that, on the basis of the correspondence
mentioned above between cavities
and solutions $\rho(r)$ to equations (\ref{einstein1}}),
we immediately have the following.
While there is 
no limit on the central density $\rho_C$ 
(or central temperature)
of a given cavity at equilibrium
(irrespective of being its size fixed or not), 
from (\ref{limitT}) on the contrary
when the size is fixed
the temperature at its boundary
(that is the temperature of the blackbody radiation
coming out through a small hole in the walls of the cavity)
is limited, whichever the central temperature is.
Fig. \ref{Tb_vs_Tc} reports 
energy density at the boundary
of the fixed-radius cavity $\rho(R)$ times $R^2$
as a function of the quantity $z$
defined above.
%%%%%%%%%%%%%%%%%%%%%%%%%%%% figure Tb_vs_Tc %%%%%%%%%%%%%%%%%
\begin{figure}\leavevmode
\begin{center}
\epsfxsize=8cm
\epsfbox{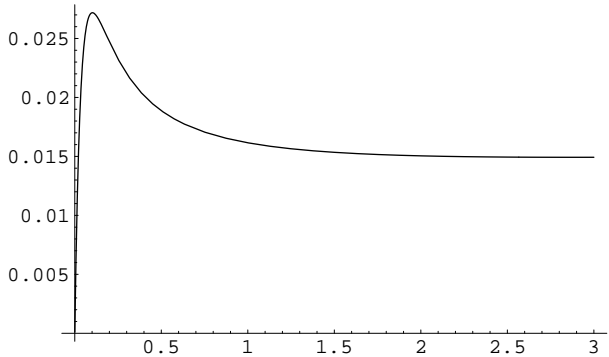}
\caption{$\rho(R) R^2$ versus $z \equiv \rho(0) R^2$ ($R$ is fixed).}
\label{Tb_vs_Tc}
\end{center}
\end{figure}
%%%%%%%%%%%%%%%%%%%%%%%%%%%%%%%%%%%%%%%%%%%%%%%%%%%%%%%
Here again we have an universal plot, not depending on $R$.
Its analytic expression in terms of the solution $f_{f(0)=1}$
corresponding to initial conditions (\ref{condinizter}) is

\begin{eqnarray}\label{tempb-tempc}
\rho(R)\ R^2 = \frac{z}{f^2_{f(0)=1}(\sqrt{8 \pi z})} 
\end{eqnarray}
as can be deduced from equation (\ref{C}) and from
the scaling property (\ref{allsolutionsf}).
This dimensionless function has absolute maximum $Q = 0.02718$
(the extremal point is $z =  0.1021$)
and goes to the asymptotic value $3/(56 \pi)$ 
(with damped oscillations around it)
when $r \rightarrow \infty$
as can be seen from (\ref{limitT}).
For a cavity of size $R$ 
we have thus the following limiting
density on the boundary

\begin{eqnarray}\label{Tmax}
\rho(R) \leq \frac{Q}{R^2}
\end{eqnarray}
The limit depends on the size $R$ of the cavity.
For temperatures at the boundary of the cavity
higher than those corresponding to this limit, 
thermodynamical equilibrium is no longer possible.
If we perform a small hole in the cavity,
the radiation emerging from it cannot have 
an energy density (or corresponding temperature or pressure)
larger than this limit.
Note that $\rho(R)$ never vanishes, whichever is
the ($\not= 0$) value of central density. 
Due to (\ref{state}) this implies
that for our cavities,
self-gravitating equilibrium configurations 
cannot exist.

We study, finally, how the radial profiles
for entropy density in the cavities depend on central density
(or temperature) and then on the mass of
the radiation inside it. As already said, 
this study is performed to see
if the results \cite{Oppen2}
on the variation of entropy radial profiles with mass
for spherically symmetric bodies with constant density
can be extended/confirmed for other spherically
symmetric equilibrium configurations.
Instead to refer to  
entropy
density in the proper orthonormal frame $\sigma(r)$,
following \cite{Oppen2}
we present our results in terms of, 
call it, entropy radial density $s(r)$
defined by

\begin{eqnarray}\label{entropy}
S = \int_{0}^{R}{4 \pi r^2 \ s(r) \ dr},
\end{eqnarray}
where $S$ is the total entropy of the cavity and $R$, as before,
its radius.
We see that $s(r)$ is the specific contribution to $S$
per unit area per radial slice $(r, r+dr)$.
Evidently the relation between $s$ and $\sigma$ is
$s(r) dr = \sigma(\ell) d\ell$, where $d\ell$ is the proper line
element, so that (from (\ref{metric}))

\begin{eqnarray}\label{entropy_density}
s = \sigma \frac{d\ell}{dr} = \sigma \sqrt{h}.
\end{eqnarray}
From first law of thermodynamics and assuming
extensivity to hold locally (from equivalence principle)
for the pertinent thermodynamical quantities,
the Gibbs-Duhem relation \cite{Gibbs}

\begin{eqnarray}\label{gibbs-duhem}
\rho = T\sigma - p +\mu n
\end{eqnarray}
obtains,
where $\mu$ is the chemical potential
and $n$ the particle (photon, in this case) number density
in the rest frame of the fluid.
Taking into accont that for photons $\mu = 0$,
from (\ref{state}) and (\ref{stefan-boltzmann}) we obtain

\begin{eqnarray}\label{entropy_density2}
s = \frac{4}{3} \frac{\rho}{T} \sqrt{h} \propto \rho^{3/4} \sqrt{h}.
\end{eqnarray}
From this equation and from scaling properties 
(\ref{allsolutionsf}-\ref{allsolutionsh}),
we have that $s$ scales as

\begin{eqnarray}\label{scaling_s}
s_\lambda(r) = \frac{1}{\lambda^{3/2}} s_1(r/\lambda),
\end{eqnarray}
where $s_1$ is the
entropy radial density for the reference configuration, 
corresponding to the functions $f_1$ and $h_1$
in equations (\ref{allsolutionsf}-\ref{allsolutionsh}).
Putting $x \equiv r/R$, with $R$ assigned,
we have

\begin{eqnarray}\label{r_over_R0}
s_\lambda(R x) = \frac{1}{\lambda^{3/2}} s_1(R x/\lambda).
\end{eqnarray}
This law express the scaling (\ref{scaling_s})
in terms of the normalized radius $x$.

\begin{figure}\leavevmode
\begin{center}
\epsfxsize=8cm
\epsfbox{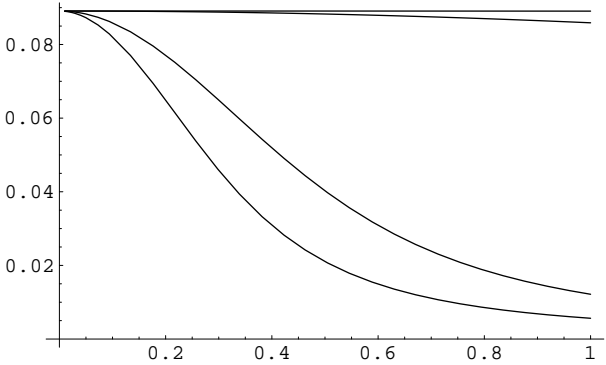}
\caption{$f(0)^{3/2} \rho^{3/4} \sqrt{h}$ versus $x \equiv r/R_0$
for various choices of central temperature, see text.
The curves go upwards at decreasing central temperature. The second
curve from bottom corresponds to the limit mass configuration.}
\label{s_vs_r}
\end{center}
\end{figure}

Let us choose as radius of the cavity
the radius $R_0$ for which $2m/r$ in Fig. \ref{2m_over_r} 
is maximum.
In order to compare entropy radial profiles
when the central density is changed,
Fig. \ref{s_vs_r} plots radial profiles of the function
$f(0)^{3/2} \rho^{3/4} \sqrt{h}$ for various choices
of $f(0)$.
%%%%%%%%%%%%%%%%%%%%%%%%%%%% figure s_vs_r %%%%%%%%%%%%%%%%%
%%%%%%%%%%%%%%%%%%%%%%%%%%%%%%%%%%%%%%%%%%%%%%%%%%%%%%%
The radial coordinate is expressed as $x = r/R_0$
and $\rho$ is intended again in units of $C$.
The curves are with increasing f(0) upwards,
respectively $f(0) = 2/3, \ 1, \ 10, \ 100$.
This means that the second curve from below
corresponds to the critical configuration,
(for which the central density $\rho(0)_{extr}$ is such that
$M$ is maximum) and the curves above and below it
are for central densities respectively
lower and higher than $\rho(0)_{extr}$.
The Figure shows that
at increasing central density
the entropy distribution is more and more
weighted around the center;
further inspection of
equation (\ref{r_over_R0}) gives 
that in the limit $\rho(0) \rightarrow \infty$
the plotted function
$f(0)^{3/2} \rho^{3/4} \sqrt{h}$ can be approximated as
$\simeq const \cdot \ [\lambda/(x R_0+\lambda)\ ]^{3/2}$,
where $\lambda = f(0)$.
Let us put together this Figure with Fig. \ref{mass_vs_T}.  
If we start from low densities (or temperatures) we see that
at increasing $M$ the contribution to total entropy
is more and more dominated by the most internal layers in $r$,
at variance with what obtained, for another system, in \cite{Oppen2} 
where at increasing $M$
the dominating role is gradually played by external layers.
We see also that
when however we are at densities a bit larger than $\rho(0)_{extr}$
the weight of external layers tends to increase with $M$,
confirming in this case what found for the system
studied in \cite{Oppen2}.
This will be reversed again after the next extremum
for $2M/R_0$ (a minimum) will be encountered and so on.
We see therefore that in our circumstances the dependence
on mass of radial entropy profile is different
and more complex than in the case examined in \cite{Oppen2}.

\section{Conclusions}
Our results can be summarized as follows.
Considering 
insulated spherical cavities 
thermodynamically at 
equilibrium, i.e. filled with blackbody radiation, 
at energy densities sufficiently high
to make not negligible the gravitational self-interaction
of radiation, we determine their mechanical equilibrium
configurations.
We find that 
equilibrium is possible without any limit
on mass $M$ (ADM mass) of the contained radiation,
provided the area radius $R$ of the
cavity is arbitrarily large. 
At fixed $R$, on the contrary,
a (very large) mass limit is present, 
which, if expressed in terms
of the Schwarzschild mass $M_S$ for that radius, is 
$M(R) \leq K M_S(R)$ with $K = 0.493$ not depending on $R$.
This implies that no more than this amount
of mass energy can be stored 
as blackbody radiation
in a cavity with assigned area radius $R$.
Furthermore we find that equilibrium is possible
for any value of central energy density (or temperature),
irrespective of being the size of the cavity
fixed or not.
The energy density $\rho_B = \rho(R)$ at the boundary of the cavity
can take any non-zero
value if the size $R$ of the cavity is
arbitrary; it is however bounded
if $R$ is fixed. We find $\rho(R) \leq Q/R^2$ 
with $Q=0.02718$ not depending on $R$.
This can be interpreted as a limit 
on the density of the radiation emerging from 
a small hole in a cavity
of assigned radius $R$.
The fact that $\rho_B$ is never 0 
(except in the trivial case for which $\rho(0) = 0$)
implies
that self-gravitating equilibrium configurations are never allowed.  
In the limit of very high central density (or temperature)
we have an asymptotic situation in which both the quantities $2M/R$
and $\rho(R) \ R^2$ 
no longer depends on $R$, 
being the first fixed to the value 3/7 
and the second to  $3/(56 \pi)$.

These results are obtained considering the radiation contained in
the cavities from a purely classical point of view,
that is neglecting any quantum process such as, for example,
pair production by photons. This has been purposely done 
to study the radial profiles
of conventional entropy density
for classical fields
under strong gravitational interaction conditions.
For central temperatures not very high
this description can be in any case safely adequate.
It turns out furthermore that when the considered radii
are roughly $\sim 100 \ R_{\odot}$ or larger,
these systems can reach even
their mass limits without any significant appearance
of pair production processes. Cavities with
$100 \ R_{\odot}$ size
(a fraction of AU) can be at equilibrium
(i.e. not collapsing) even with masses
of orders of $10^7 M_{\odot}$.

\end{document}